\documentclass[twocolumn,showpacs,preprintnumbers,amsmath,amssymb,footinbib]{revtex4}

\usepackage{graphicx}
\usepackage{dcolumn}
\usepackage{bm}
\usepackage{epsfig}
\usepackage{epstopdf}
\usepackage{enumerate}
\usepackage{mathrsfs}

\begin{document}


\title{Theory of cavity-enhanced spontaneous four wave mixing}

\author{K. Garay-Palmett$^1$ Y. Jeronimo-Moreno$^2$, and A. B. U'Ren$^1$}

\affiliation{$^1$Instituto de Ciencias Nucleares, Universidad Nacional Aut\'onoma de M\'exico, apdo. postal 70-543, M\'exico 04510 DF
\\$^2$Instituto de F\'isica, Universidad Nacional Aut\'{o}noma de M\'{e}xico, apdo. postal 70-543, M\'exico 04510 DF}

\newcommand{\epsfg}[2]{\centerline{\scalebox{#2}{\epsfbox{#1}}}}

\begin{abstract}
In this paper we study the generation of photon pairs through the process of spontaneous four wave mixing (SFWM) in a $\chi^{(3)}$ cavity.  Our key interest is the generation of photon pairs in a guided-wave configuration - fiber or waveguide -  where at least one of the photons in a given pair is matched in frequency and bandwidth to a particular atomic transition, as required for the implementation of photon-atom interfaces.   We present expressions, along with plots, for the two-photon joint intensity both in the spectral and temporal domains.   We also present expressions for the absolute brightness, along with numerical simulations, and show that the presence of the cavity can result in a flux enhancement relative to an equivalent source without a cavity.
\end{abstract}

\maketitle


\section{Introduction}

Photon pair generation can be accomplished through spontaneous parametric processes including spontaneous parametric downconversion (SPDC), based on a second-order non-linearity, and spontaneous four wave mixing (SFWM), based on a third-order non linearity.  Both of these processes are extremely versatile in terms of the resulting properties which emitted photon pairs can have.  Thus, SFWM has been used, or proposed, as the basis for the generation of factorable photon pairs~\cite{clark11,soller11,soller10,halder09,cohen09,garay07}, of photon pairs with arbitrary spectral correlation properties~\cite{fan07,rarity05,fan05}, of polarization-entangled photon pairs~\cite{medic10,fan07a,lee06}, and of ultrabroadband photon pairs~\cite{garay08}.  In addition, if implemented in a fiber, SFWM has a number of important advantages including the possibility of very long interaction lengths and straightforward integration with existing fiber optic networks. Recently, in the context of classical optics have been important experimental demonstrations of four-wave mixing (FWM) \cite{Ivanin10}, and the implementation of nonlinear optical devices, such as wavelength converters and multi-wavelength fiber lasers which are based on FWM \cite{Zamzuri11,Ahmad11,Chen10,Sun10,Chen11}. While the use of a guided wave configuration constrains the transverse spatial properties of emitted photon pairs, the spectral properties can vary widely.   In common with SPDC, in the process of SFWM the energy splitting ratio in a given photon pair is governed only by phasematching, so that while the emission bandwidth can vary greatly in accordance with the properties of the nonlinear material and the pump, this bandwidth cannot typically be made naturally very small.   In this paper we explore the use of $\chi^{(3)}$ cavities for the generation of SFWM photon pairs which are sufficiently narrowband for the effective implementation of atom-photon interfaces, i.e. so that one photon from a SFWM photon pair may be reliably absorbed by a singe atom.  This is an important line of research which represents a route for the implementation of quantum memories~\cite{reim10,lvovsky09}.

This paper extends to the process of SFWM a previous analysis from our group, see~\cite{jeronimo10}, of cavity-enhanced SPDC~\cite{Ou1999,raymer05,Kuklewicz2006,Neergaard-Nielsen2007,Bao2008,Wang2008,Haase2009,Scholz2009,scholz,Wang2010,Zhang2011,Pomarico2011}.  In our earlier paper we analyzed the spectral and temporal properties of SPDC photon pairs produced in a nonlinear cavity.  We showed that the use of a high-finesse cavity results in photon pair emission constrained to the narrow spectral windows defined by the cavity modes.  In addition we studied how the spectral mode structure imposed by the cavity translates into a specific mode structure in the temporal domain, and we showed that for a sufficiently narrowband pump, the presence of a cavity leads to a flux enhancement relative to an equivalent source without a cavity.  In the present paper, we have carried out a corresponding analysis for SFWM.  Because resonance at the emission frequencies only (i.e. excluding the pump) yields both narrowband emission modes and a cavity-induced flux enhancement, in this paper we concentrate on this case which is technologically simpler to implement.  In addition, in this paper we separately  analyze the cases of resonance to one, or both of the emission modes.   We discuss the presence of high-dimensional temporal entanglement in SFWM photon pairs emitted in a $\chi^{(3)}$ cavity, and we present an analytic expression, in closed form, for the joint temporal intensity valid under certain approximations. We present expressions for the absolute brightness along with corresponding numerical simulations, as well as a simple geometrical model which leads to an improved understanding of the cavity-induced flux enhancement.

A $\chi^{(3)}$ SFWM cavity source can be implemented with an optical fiber on which two Bragg gratings are recorded to form a cavity [see Fig.~\ref{Fig:esquema}(a)].  Likewise, the cavity can be in the form of a fiber ring  \cite{Zamzuri11,Ahmad11,Chen10}, which could be implemented through a length of fiber connected to one input port and one output port of a fiber-based beam splitter and two lengths of fiber, to form the input and output of the device, connected to the remaining two ports [see Fig.~\ref{Fig:esquema}(b)].  Alternatively,  recent work has shown the feasibility of observing optical non-linear effects in waveguide micro-rings, where light is coupled and out-coupled from the cavity through a straight waveguide placed tangentially to the ring~\cite{clemmen11}.  This latter route represents a highly promising approach, which could be used for generating SFWM photon pairs and which could be incorporated into integrated optics circuits.  This paper aims to present the required theory for the description of various aspects of SFWM cavity sources.

\section{The SFWM two-photon quantum state}
\label{Sec:STATE}

In this paper we are interested in studying the spectral and temporal properties of photon pairs generated by spontaneous four wave mixing (SFWM) in a fiber-based (or waveguide-based) optical cavity.   Let us begin by reviewing the SFWM two-photon state produced by a fiber, in the absence of a cavity, assuming that each of the participating modes propagates in a single transverse mode.  The state which describes the signal(s) and idler(i) state produced by SFWM is given by $|\Psi\rangle=|0\rangle_s|0\rangle_i +\zeta|\Psi_2\rangle$, in terms of the vacuum $|0\rangle_s|0\rangle_i$ and the two-photon state $|\Psi_2\rangle$, where $\zeta$ is a constant related to the conversion efficiency.    We have shown previously that $|\Psi_2\rangle$ is given by~\cite{garay07,garay10}

\begin{equation}
\label{state2}|\Psi_2\rangle=\sum_{k_s}\sum_{k_i}G(k_s,k_i)
\hat{a}^{\dag}(k_s)\hat{a}^{\dag}(k_i)|0\rangle_s|0\rangle_i,
\end{equation}

\noindent where $\hat{a}(k_\mu)$ (with $\mu=s,i$) represents the annihilation operators for each of the signal and idler modes, labeled by the wavenumber $k_\mu$. Expressed in terms of frequencies, the function $G(\omega_s,\omega_i)$ is the joint spectral amplitude (JSA) given by $G(\omega_s,\omega_i)=\ell(\omega_s)\ell(\omega_i)F(\omega_s,\omega_i)$ in terms of $\ell(\omega)=\sqrt{\hbar \omega/[\pi\epsilon_o n^2(\omega)]}$ and in terms of the function

\begin{eqnarray}
\label{jsa}F(\omega_s,\omega_i)=\int\!\!&d\omega&\alpha(\omega)\alpha(\omega_s+\omega_i-\omega)\nonumber\\
&\times&\mbox{sinc}\!\left[\frac{L\Delta
k (\omega_s,\omega_i,\omega)}{2}\right]e^{i\frac{L\Delta k(\omega_s,\omega_i,\omega)}{2}}.
\end{eqnarray}

In Eq.~(\ref{jsa}), $L$ is the fiber length, $\alpha(\omega)$ represents the pump spectral
envelope function and $\Delta k(\omega_s,\omega_i,\omega)$ is the phase
mismatch

\begin{eqnarray}
\label{delk}\Delta
k(\omega_s,\omega_i,\omega)&=k(\omega)+k(\omega_s+\omega_i-\omega)-k(\omega_s)
\nonumber\\&-k(\omega_i)-2\gamma P,
\end{eqnarray}

\noindent where $\gamma$ and $P$ are the nonlinear
coefficient (related to self-phase and cross-phase modulation
effects) and the peak pump power, respectively.

Assuming that the pump has a Gaussian spectral profile, with bandwidth
$\sigma$, the coefficient
$\zeta$ in Eq.~(\ref{state2}) is given by

\begin{equation}
\label{zeta}{\zeta}=i\frac{2(2\pi)^{1/2}\epsilon_0cn(\omega_{o})}{\hbar
\omega_{o}}\frac{\gamma_{fwm}
LP}{\sigma}\delta k,
\end{equation}

\noindent where $\epsilon_0$ is the permittivity of free space, $c$ is the speed of light, $\omega_{ o}$ is the central frequency for the pump, and $n(\omega)$ is the refractive index of the fiber. The term $\gamma_{fwm}$ is the nonlinear coefficient associated with the SFWM
interaction, and $\delta k$ is the mode spacing obtained in the quantization of the electromagnetic field.

\subsection{Spectral-domain description of a SFWM cavity source}

\begin{figure}[ht]
\centering\includegraphics[width=0.4\textwidth]{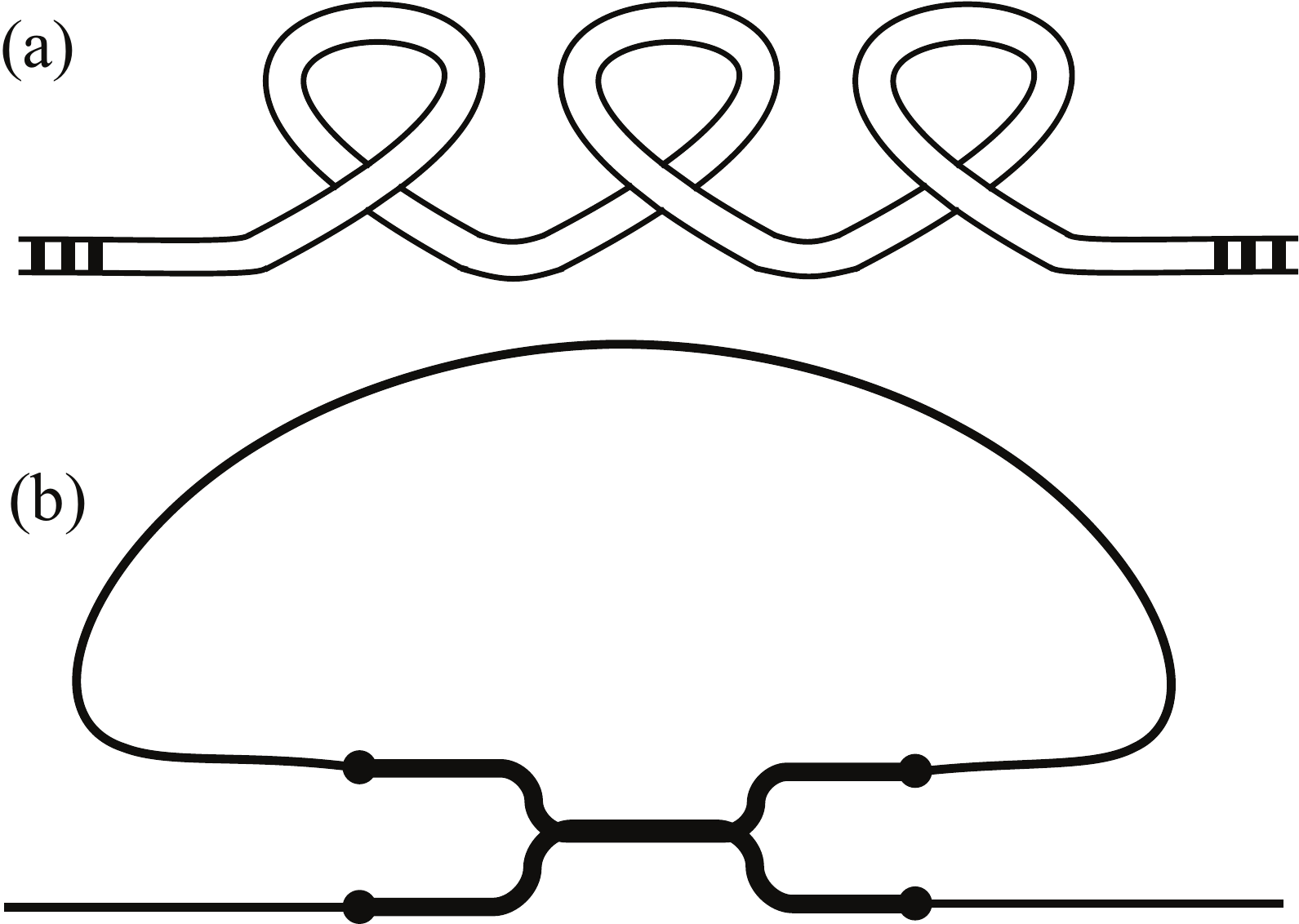}\caption{(color online) Two ways in which a SFWM cavity source could
be implemented.  In (a), two Bragg mirrors are written on a fiber, thus forming a cavity.  In (b), a ring cavity (fiber- or waveguide-based) is used.}
 \label{Fig:esquema}
\end{figure}

Let us now turn to the case where the SFWM fiber is in the form of a cavity.  This could be done in one of two ways: i) Two Bragg mirrors with a separation $L$ could be written on a length of fiber thus forming a cavity, or  ii) A fiber or waveguide ring cavity, of circulation length $L_c$ could be used, coupled to a straight fiber or waveguide placed tangentially to the ring.   In this paper we will study cases where the cavity used is resonant to one or both of the SFWM modes, while it is not resonant to the pump frequency. Fig.~\ref{Fig:esquema} shows, schematically, both of these types of fiber cavity sources.  While we are interested in both the pulsed and continuous-wave pump regimes, we will carry out our analysis for the pulsed case and will obtain the continuous-wave case as an appropriate limit.

For simplicity, in this paper we restrict our attention to the SFWM process with degenerate pumps, i.e. for which both pump photons in a given event come from the same pump mode.  In our analysis below, we will assume a cavity formed by two mirrors, although we will indicate how our results would apply to the case of a ring cavity.  Because in our calculation the cavity is not resonant to the pump, as mentioned above, both mirrors are perfectly transmissive for the pump.  We will refer to a fiber cavity SFWM source resonant only to the signal-photon as Cs cavity, to a source resonant only to the idler-photon as a Ci cavity, and to a source resonant to both SFWM photons as a Csi cavity.  We also assume that the left-hand cavity mirror (through which the pump enters, to be referred to as mirror $1$) is perfectly reflective for SFWM photons, so that all
emitted photons are forward-propagating, transmitted by the
right-hand cavity mirror, to be referred to as mirror $2$.  The calculation below, closely follows that for SPDC reported by us in an earlier paper, see Ref.~\cite{jeronimo10}.

For the Csi cavity it can be shown that the two-photon component of the state at the cavity output becomes

\begin{equation}
\label{state3}|\Psi_2^{si}\rangle=\zeta\sum_{k_s}\sum_{k_i}G_{si}(k_s,k_i)a_s^{\dag}(k_s)
a_i^{\dag}(k_i)|0\rangle_s|0\rangle_i,
\end{equation}

\noindent where we have defined the joint amplitude function

\begin{equation}
\label{JSACav}G_{si}(\omega_s,\omega_i)=G[k_s(\omega_s),k_i(\omega_i)]A_{s}(\omega_s)A_{i}(\omega_i),
\end{equation}

\noindent given in terms of $G(k_s,k_i)$ [see
Eq.~(\ref{jsa})], but expressed here in terms of
frequencies, and where the function $A_{\mu}(\omega)$
describes the effect of the cavity and can be written as

\begin{equation}
\label{E:A}A_{\mu}(\omega_{\mu})=\frac{t_{2\mu}}{1-|r_{2\mu}
|e^{i(2\beta_{\mu}+\delta_{1\mu}+\delta_{2\mu})}}.
\end{equation}

In Eq.~(\ref{E:A}), $t_{2\mu}$ and  $r_{2\mu}=|r_{2\mu}|e^{i\delta_{2\mu}}$ are the amplitude transmissivity and the amplitude reflectivity of the right-hand cavity mirror, respectively, while $r_{1\mu}=e^{i\delta_{1\mu}}$ is the amplitude reflectivity of the left-hand cavity mirror, for
$\mu=s,i$.  In Eq.~(\ref{E:A}), $\beta_{\mu}=k(\omega_{\mu})L$ represents the phase acquired by
photon $\mu$ in one round-trip of the fiber cavity.

The cavity-modified joint spectral
amplitude function is shown in Eq.~(\ref{JSACav}). The corresponding cavity-modified joint spectral
intensity (JSI) $S_{si}(\omega_s,\omega_i)=|G_{si}(\omega_s,\omega_i)|^2$ is given by~\cite{jeronimo10}

\begin{equation}
\label{JSIcav}S_{si}(\omega_s,\omega_i)=|F(\omega_s,\omega_i)|^2\mathscr{A}_{s}(\omega_s)
\mathscr{A}_i(\omega_i),
\end{equation}

\noindent with $F(\omega_s,\omega_i)$ given by Eq.~(\ref{jsa}), and

\begin{equation}
\label{airy}\mathscr{A}_{\mu}(\omega_{\mu})=\frac{|t_{2\mu}|^2}
{\left(1-|r_{2\mu}|\right)^2}\frac{1}{1+\mathscr{F}_{\mu}\sin^2[\Delta_{\mu}(\omega_{\mu})/2]},
\end{equation}

\noindent written in terms of the coefficient of finesse $\mathscr{F}_{\mu}$

\begin{equation}
\label{fine}
\mathscr{F}_{\mu}=\frac{4|r_{2\mu}|}{\left(1-|r_{2\mu}|\right)^2},
\end{equation}

\noindent and the phase factor

\begin{equation}
\label{delt}
\Delta_{\mu}(\omega_{\mu})=2\beta+\delta_{1\mu}+\delta_{2\mu}.
\end{equation}

Note that in the case of a ring cavity, mirror $1$ does
not exist, while mirror $2$ becomes a fiber-based beam
splitter as shown in Fig.~\ref{Fig:esquema}(b).  The coefficient $r_2$ is then
the amplitude associated with an intra-cavity
photon remaining in the cavity in a given iteration, while the
coefficient $t_2$ is the amplitude associated with an
intra-cavity photon exiting the cavity in a given iteration.
Because mirror $1$ does not exist,
$\delta_{1\mu}$ should be replaced with $0$.  Also, the fiber length $L$ should be
replaced with $L_r/2$, where $L_r$ is the circulation length in the ring cavity.

The Airy function $\mathscr{A}_{\mu}(\omega_{\mu})$
is formed by a sequence of equal height peaks, with width $\delta\omega_\mu$, and
spectral separation, or free spectral range, $\Delta\omega_\mu$. Under the assumption that the
index of refraction remains constant among different peaks, it can
shown that \cite{jeronimo10}

\begin{equation}
\label{sep} \Delta\omega_\mu=\frac{\pi c}{L
n_\mu},
\end{equation}

\noindent and

\begin{equation}
\label{width} \delta\omega_\mu=\frac{2c}{L
n_\mu \sqrt{\mathscr{F}_{\mu}}}=\frac{2 \Delta \omega}{\pi \sqrt{\mathscr{F}_\mu}},
\end{equation}

\noindent where $n_\mu=n(\omega_{\mu0})$ is the refractive index of
the fiber evaluated at the central frequency of the signal
and idler modes ($\mu=s,i$). Note that the cavity is resonant at
frequencies at which the peaks of the Airy function appear. Thus, in
order to ensure resonance at a particular frequency $\omega$,
the condition $\sin[\Delta_\mu(\omega)/2]=0$ must be fulfilled.
Also note that resonance at a given desired frequency can be achieved by
adjusting the reflection phases $\delta_{1\mu}$ and $\delta_{2\mu}$.

For the case of a cavity resonant at only one of the SFWM modes, say the
signal mode, the two-photon state and the joint spectral intensity are given by Eqs.~(\ref{state3}) and (\ref{JSIcav}), respectively, with $|r_{2i}|=0$. Thus, $\mathscr{F}_{i}=0$ and $\mathscr{A}_{i}(\omega_{i})=1$, and consequently the JSI for the Cs cavity becomes

\begin{equation}
S_{s}(\omega_s,\omega_i)=|F(\omega_s,\omega_i)|^2\mathscr{A}_{s}(\omega_s).
\end{equation}

Likewise, for a Ci cavity the JSI becomes

\begin{equation}
S_{i}(\omega_s,\omega_i)=|F(\omega_s,\omega_i)|^2\mathscr{A}_{i}(\omega_i).
\end{equation}

\begin{figure}[ht]
\centering\includegraphics[width=0.43\textwidth]{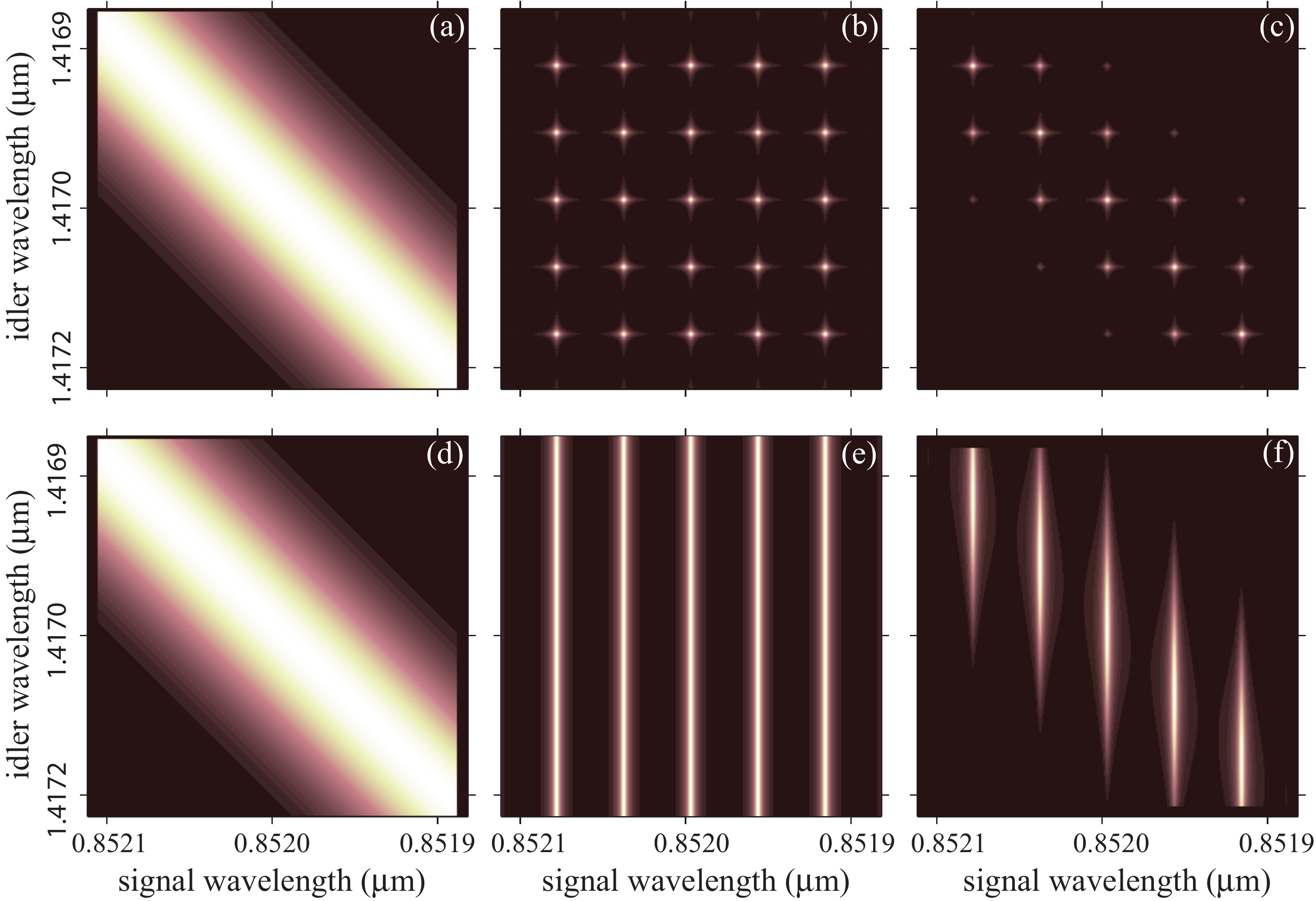}\caption{(color online) \label{Fig:JSI}
In this figure we show the joint spectral intensity for SFWM photon pairs produced in a $\chi^{(3)}$ cavity.   In the first three panels we present, for a Csi cavity,  plots prepared for a particular set of parameters (see main text), as a function of the signal and idler frequencies, of: (a) the joint spectral intensity in the absence of a cavity, (b) the contribution of the cavity, comprised of a matrix of cavity-allowed modes, and (c) the resulting cavity-modified joint spectral intensity.  Panels (d)-(f) are similar to panels (a)-(c), except that resonance at the idler mode has been suppressed (i.e. this corresponds to the Cs cavity).}
\end{figure}

The spectral structure of a SFWM two-photon state obtained with a Csi cavity is illustrated in Figs.~\ref{Fig:JSI}(a)-(c), for which we have assumed a photonic crystal fiber (PCF) with $0.68\,\mu$m core radius, and $0.5$ air-filling fraction, and we have relied on the step index model~\cite{Wong2005}.  These parameters have been chosen to achieve phasematching for a SFWM process with degenerate pumps centered at $\omega_p=2 \pi c/1.064\mu $m and signal an idler modes centered at $\omega_s^o=2 \pi c/0.852\mu $m and $\omega_i^o=2 \pi c/1.417\mu $m, respectively.  The SFWM frequencies are chosen to be considerably frequency non-degenerate, which facilitates photon-pair splitting.  We have assumed a fiber length of $L=1$cm and a pump bandwidth of $\sigma=80$ GHz. Fig.~\ref{Fig:JSI}(a) represents the JSI without cavity, i.e. so that  $\mathscr{A}_{s}(\omega)=\mathscr{A}_{i}(\omega)=1$. The two-dimensional Airy pattern describing the effect of the cavity is shown in Fig.~\ref{Fig:JSI}(b), while Fig.~\ref{Fig:JSI}(c) represents the resulting JSI for the cavity-modified SFWM photon pairs.  While for illustration purposes, here we have assumed $r_2=0.8$, corresponding to $\mathscr{F}=80$,  in a realistic implementation the finesse would probably be chosen much higher to ensure the emission of sufficiently narrowband photon pairs.
Note that for these plots [\ref{Fig:JSI}(a)-(c)], we have assumed that the signal and idler photons are filtered using rectangular filters of width $\sigma_{f \mu}=5\Delta\omega_\mu$ (with $\mu=s,i$), centered at $\omega_s^o$ and $\omega_i^o$, respectively.

As can be noted, the effect of the cavity is that the emission frequencies are re-distributed, with respect to the corresponding emission frequencies in the absence of a cavity, so that emission occurs only for frequencies within the cavity modes. Because the mode width $\delta \omega$ scales as $\mathscr{F}^{-1/2}$,  cavity-enhanced SFWM with a sufficiently high finesse can be an effective route towards narrowband photon pairs, to be used for example in the context of atom-photon interfaces. As will be studied further in a later section, if the pump bandwidth is of the order of the cavity mode width $\delta \omega$, a significant enhancement to the source brightness can result with respect to an equivalent source without a cavity.

In Figs.~\ref{Fig:JSI}(d)-(f) we illustrate the spectral structure of SFWM photon pairs produced by a Cs cavity.  Here we assume an identical source configuration to that assumed for the Csi cavity, except that the resonance for the idler photon has been suppressed.  Fig.~\ref{Fig:JSI}(d) represents the JSI without cavity, i.e. so that  $\mathscr{A}_{s}(\omega)=\mathscr{A}_{i}(\omega)=1$. The one-dimensional Airy pattern describing the effect of the cavity resonant for the single-mode only is shown in Fig.~\ref{Fig:JSI}(e), while Fig.~\ref{Fig:JSI}(f) represents the resulting JSI for the cavity-modified SFWM photon pairs.   As for the Csi cavity, we have assumed that the signal and idler photons are filtered using rectangular filters with width $\sigma_{f \mu}=5\Delta\omega_\mu$, centered at $\omega_s^o$ and $\omega_i^o$, respectively.

As may be appreciated, the effect of the cavity is a spectral re-distribution for the resonant, in this case signal, photon.  The result is that the JSI is now formed by elongated modes of width $\delta \omega$ for the signal photon and a larger width determined by the cavity length and pump bandwidth for the idler photon.

\subsection{Temporal-domain description of a SFWM cavity source}

It is instructive to study the two-photon state in the temporal domain, in addition to the spectral domain used in our description so far.  To this end, we define the joint temporal amplitude (JTA) $\tilde{f}(t_s,t_i)$ as the two-dimensional Fourier transform of the JSA, i.e.

\begin{equation}
\tilde{f}(t_s,t_i)=\mathcal{F}\{G_{si}(\omega_s,\omega_i)\},\label{E:JTI}
\end{equation}

\noindent where $\mathcal{F}$ denotes Fourier transform. and $G_{si}(\omega_s,\omega_i)$ is given according Eq.~(\ref{JSACav}).

The joint temporal intensity (JTI) can now be written as $S_t(t_s,t_i)=|\tilde{f}(t_s,t_i)|^2$, which when properly normalized yields the time of emission joint probability distribution.

\begin{figure}[ht]
\centering\includegraphics[width=0.4\textwidth]{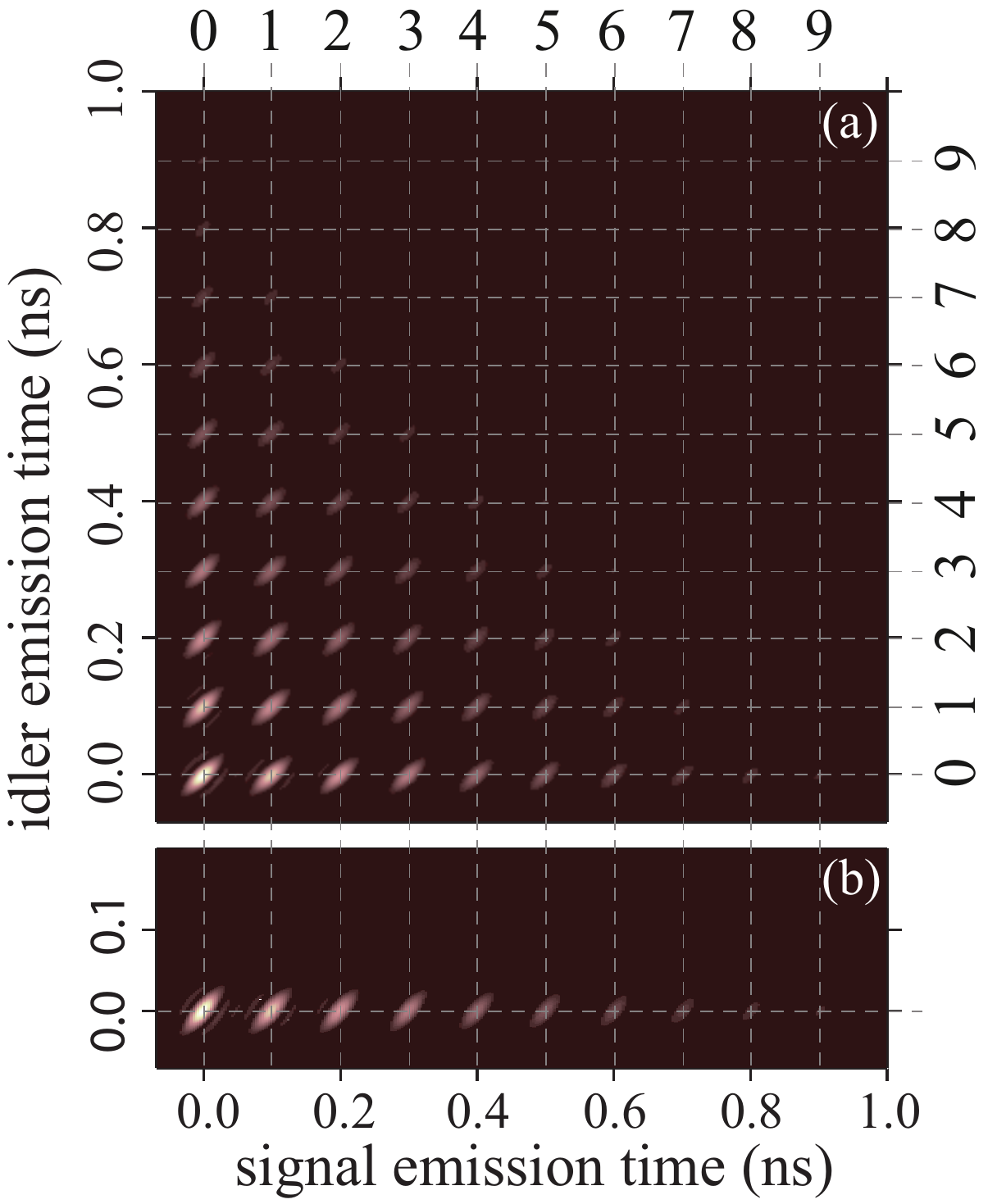}\caption{\label{Fig:JTIsinrotar} (a) Here we show for the same parameters as in Fig.~\ref{Fig:JSI} the joint temporal intensity for the Csi cavity, plotted  as a function of the signal and idler times of emission.  The vertical (horizontal) dotted lines indicate possible signal-mode (idler-mode) times of emission, each corresponding to an integer number of cavity round-trips.  (b) This panel is similar to panel (a), except that it has been prepared for the Cs cavity.}
 \end{figure}

The spectral cavity mode structure translates into a specific temporal mode structure.  In Fig.~\ref{Fig:JTIsinrotar}, we have shown for the same source parameters as in Fig.~\ref{Fig:JSI} the resulting JTI calculated numerically through Eq.~(\ref{E:JTI}). In physical terms, the two photons in a given pair may exit the cavity together, or may be delayed with respect to each other by a whole number of cavity round-trip times.  The vertical (horizontal) dotted lines indicate possible signal-mode (idler-mode) emission times, corresponding to a whole number of cavity round-trip times.   This results in a matrix of temporal emission modes, as is apparent in Fig.~\ref{Fig:JTIsinrotar}, which decay in amplitude for increasing times of emission (which imply an increasing number of cavity round trips).  Of course, for a higher finesse, this amplitude roll-off is slower resulting in the appearance of a larger number of temporal modes.

Let us label each of the emission modes according to the cavity iteration at which each of the two photons is emitted, so that the state $|ij\rangle$, with a corresponding probability amplitude $C_{ij}$, means the signal photon emitted in the $i$th cavity iteration and the idler photon emitted in the $j$th cavity iteration.   We may then write down the resulting two-photon state as follows

\begin{align}
|\Psi\rangle&=[C_{00}|00\rangle]+[C_{10} |10\rangle+C_{01} |01\rangle] \nonumber\\
&+ [C_{20}|20\rangle+C_{11} |11\rangle +C_{02} |02\rangle]  \nonumber\\
&+ [C_{30}|30\rangle+C_{21}|21\rangle+C_{12}|12\rangle+C_{03}|03\rangle]+....\label{E:ent}
\end{align}

Here, each group of terms in square brackets indicates a certain fixed number of total cavity iterations.  Thus, the only term in the first square-bracket corresponds to both photons emitted together without any intra-cavity reflections.  The two terms in the second square-bracket correspond to a total number of $1$ cavity iterations between the two photons.  The three terms in the third square-bracket correspond to a total number of $2$ cavity iterations between the two photons, and so forth for higher-order terms.  It can be
appreciated from Fig.~\ref{Fig:JTIsinrotar}(a) that all modes within a given square-bracket group involve the same amplitude.

Note that the state in Eq.~(\ref{E:ent}) exhibits quantum entanglement in the temporal mode degree of freedom.  Note also that the state of each photon is described by a Hilbert space with a dimension of the order of the number of effective cavity round trip times before the state is extinguished.  Thus, for a higher coefficient of finesse, there will be more cavity round trip times, and the dimension of the Hilbert space which describes each photon will increase.  The triangular shape of the overall pattern of modes [see Fig.~\ref{Fig:JTIsinrotar}(a)] implies that the state is non-factorable.  Thus, interestingly, this physical system leads to higher-dimensional entanglement in the temporal-mode degree of freedom with a dimension which can be tailored by adjusting the cavity finesse.

We can carry out a similar analysis for the Cs cavity.  The resulting joint temporal intensity is plotted in Fig.~\ref{Fig:JTIsinrotar}(b).  The fact that only one photon is now resonant, implies that the temporal mode structure now appears only for the signal photon.  Note that this state may now be written as $|\Psi\rangle=C_{00}|00\rangle+C_{01}|01\rangle+C_{02}|02\rangle+...$, which is factorable; hence, this system does not lead to entanglement in the temporal-mode degree of freedom.

\begin{figure}[ht]
\centering\includegraphics[width=0.4\textwidth]{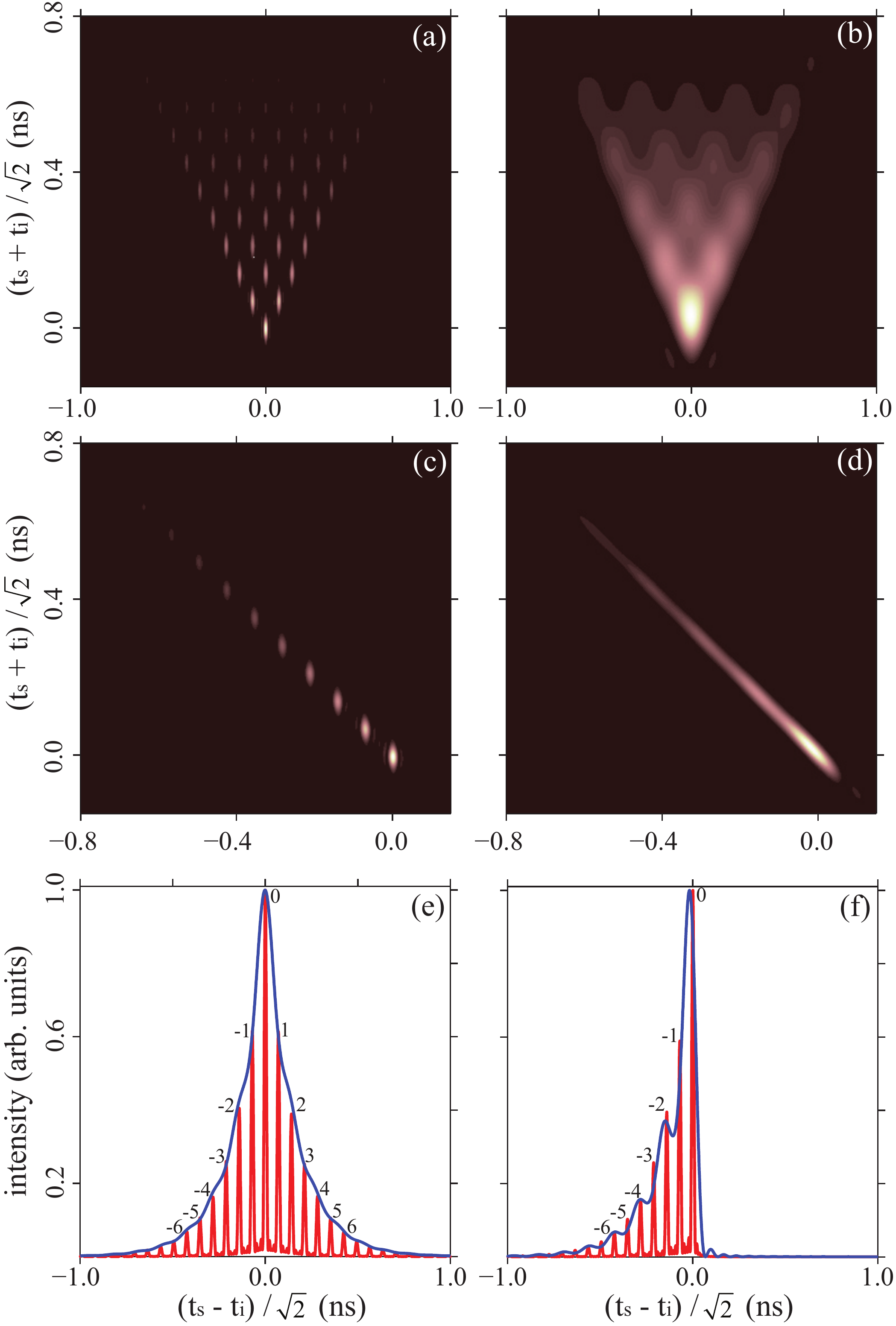}\caption{For the same parameters as in Figs.~\ref{Fig:JSI} and \ref{Fig:JTIsinrotar}, Panel (a) shows the joint temporal intensity, for the Csi cavity,  plotted as a function of the time-sum $t_s+t_i$ and time difference $t_s-t_i$ variables. Panel (b) is similar to (a) except that we have assumed that the signal and idler photons are filtered in such a way that a single spectral cavity mode is retained.  (e) represents the time-difference marginal joint temporal intensity, obtained from integrating the joint temporal intensity over the time-sum variable. Note that while the red curve (with distinct peaks) corresponds to panel (a), the blue curve corresponds to panel (b).  Panels (c)-(d) and (e) are similar to panels (a)-(b) and (e), except that the idler-mode resonance has been suppressed (i.e. this corresponds to a Cs cavity).\label{Fig:JTI}}
\end{figure}

In order to gain further physical insight into the nature of the two-photon state in the temporal domain, in Fig.~\ref{Fig:JTI}(a) we plot the joint temporal intensity for the same Csi cavity source, this time as a function of the time sum $t_s+t_i$ and time difference $t_s-t_i$ variables.  Fig.~\ref{Fig:JTI}(b) is similar to the previous plot, except that the state has been spectrally filtered so that only the central spectral mode in the JSI is retained.  It may be seen from these plots that the triangular pattern which appears in the latter JTI may be thought of as an ``envelope'' for the JTI corresponding to the full, i.e. spectrally-unfiltered, state.  It is interesting to consider the time of arrival difference distribution, which can be obtained from integrating the JTI in panels (a) and (b) over the time sum variable.   The result is plotted in Fig.~\ref{Fig:JTI}(e), where the curve composed of multiple peaks corresponds to the spectrally-unfiltered state and the other curve which again may be thought of as an envelope corresponds to the central peak of the JSI.  The multiple peaks visible in the unfiltered time of arrival difference distribution are labeled as $0$,$\pm 1$,$\pm 2$... so that a value $j>0$ means that the signal photon is emitted $j$ cavity iterations \textit{after} the idler photon, wheras a value $j<0$ means that the signal photon is emitted $j$ cavity iterations \textit{before} the idler photon.  The spacing between peaks corresponds to the cavity round trip time.

Figs.~\ref{Fig:JTI}(c), (d) and (f) are similar to panels (a)-(b) and (e), except that they correspond to the Cs cavity source (where the two sources are otherwise identical to each other).  Note that in this case the time of arrival difference distribution is asymmetric, where positive values do not occur, i.e. the signal photon in a given pair is always are emitted after the corresponding  idler photon.

\subsection{Analytical expression in closed form for the joint temporal intensity }

As was shown in the previous section, we can obtain the joint temporal intensity by numerical evaluation of Eq.~(\ref{E:JTI}).  However, in this section we show that it is possible to obtain an expression in closed analytic form under certain approximations.  These approximations include: i) describing each cavity mode through a Gaussian function of the signal and idler frequencies, ii) assuming that the SFWM mode widths are equal, i.e. $\delta \omega_s=\delta \omega_i\equiv \delta \omega$, and that the mode spacings are likewise equal, i.e. $\Delta\omega_s=\Delta\omega_i\equiv \Delta\omega$ , and iii) assuming that the joint spectral amplitude can be described only in terms of the pump bandwidth and the cavity mode structure, so that the sinc function in Eq.~(\ref{jsa}) can be replaced by unity.  Also, let us assume that each of the signal and idler modes is filtered with rectangular spectral filters of width $M \Delta\omega$, centered around the particular cavity mode of interest.  Thus, we can write the JSA in the form

\begin{align}
\Upsilon(\nu_s,\nu_i)&=e^{-\frac{(\nu_s+\nu_i)^2}{2\sigma^2}}\sum_{l=-M}^{M}\sum_{m=-M}^{M}
e^{-\frac{(\nu_s-l\Delta\omega)^2}{\delta\omega^2}} \nonumber \\
&\times e^{-\frac{(\nu_i-m\Delta\omega)^2}{\delta\omega^2}},
\label{E:jsaAprox}
\end{align}

\noindent which has been expressed in terms of the frequency detunings $\nu_{\mu}=\omega_{\mu}-\omega_{\mu 0}$ (with $\mu=s,i$).  The joint temporal amplitude, 
$\tilde{\Upsilon}(t_s,t_i)$, can now be calculated analytically as the Fourier transform of $\Upsilon(\nu_s,\nu_i)$.   It is convenient to express the joint temporal intensity in terms of the time-sum $t_{+}=(t_s+t_i)/\sqrt{2}$ and time-difference $t_{-}=(t_s-t_i)/\sqrt{2}$ variables, obtaining

\begin{align}
\tilde{S}_t(t_-,t_+)&= \exp\left(-\frac{t_-^2}{\tau_c^2} -\frac{t_+^2}{\tau^2} \right) \nonumber\\
&\times\frac{\sin^2\{(M+\frac{1}{2})\frac{\Delta\omega}{\sqrt{2}}[t_- -\left(\frac{\tau_c}{\tau}\right)^2t_+]\}}{\sin^2\{\frac{\Delta\omega}{2 \sqrt{2}}[t_- -\left(\frac{\tau_c}{\tau}\right)^2t_+]\}}   \nonumber\\
&\times\frac{\sin^2\{(M+\frac{1}{2})\frac{\Delta\omega}{\sqrt{2}}[t_- +\left(\frac{\tau_c}{\tau}\right)^2t_+]\}}{\sin^2\{\frac{\Delta\omega}{2\sqrt{2}}[t_- +\left(\frac{\tau_c}{\tau}\right)^2t_+]\}} ,  \label{E:jtiAprox}
\end{align}

In Eq.~\ref{E:jtiAprox}, $\tau_c=\sqrt{2}/\delta\omega$ is the temporal width along the $t_-$ direction, or correlation time, which defines the uncertainty in the time of emission difference. Note that this parameter is proportional to $\sqrt{\mathscr{F}}$ [see Eq.~(\ref{width})], so that increasing the finesse leads to more signal and idler cavity round-trips and hence to a greater correlation time. Parameter $\tau$ in Eq.~\ref{E:jtiAprox}, given by $\tau=\tau_c\sqrt{\delta\omega^2+\sigma^2}/\sigma$, represents the temporal width along the $t_+$ direction.

\section{A specific source design}
\label{Sec:SpDesign}

In this section we wish to present a specific design of a SFWM
photon-pair cavity source to be used as the basis for an
atom-photon interface.  Thus, we wish to match the frequency and
the emission bandwidth of one of the two SFWM modes, say
the signal mode, to an atomic transition.  In this
example, we will consider as a specific example of an atomic transition the D2 line in cesium which occurs at
$852$nm with a bandwidth of $2 \pi\times5.22$ MHz.

Note that appropriate matching of one of signal SFWM modes to the atomic transition
can be accomplished through a Cs cavity, which is
resonant only for the signal mode.  However, as will be
studied in the next section, an important advantage of the Csi cavity vs
the Cs (or Ci) cavity, is that it leads to a flux enhancement with respect
to the flux attainable in an equivalent source without a cavity.   Thus,
for the example to be shown here, we will assume that both the signal
and idler modes are resonant in the cavity, in other words we will assume that our source is based on
a Csi cavity.

As a first step in the design of a suitable SFWM cavity source,  a particular
fiber geometry must be determined which fulfils phasematching with appropriate pump and idler
frequencies so that the signal frequency is precisely matched to the atomic transition.
Relying on the step-index model~\cite{Wong2005} for the description
of the dispersion in a photonic crystal fiber (PCF), we have found that
a PCF with core radius $r=0.68\mu$m, and air-filling fraction $f=0.5$,
exhibits phasematching for degenerate pumps at $\lambda_p=1.064\mu$m, with
SFWM wavelengths $\lambda_s=0.852\mu$m
and $\lambda_i=1.417\mu$m.  Note that the $\lambda_s$ value is matched to the atomic
transition. Note also
that if the PCF geometry were to be taken into account fully, this would
lead to the need for a slight adjustment to the fiber specifications, for our choice of
frequencies.

The detection of an idler photon at $\lambda_i=1.417\mu$m could herald
the presence of a single photon in the signal mode, at $\lambda_s=0.852\mu$m,
to interact with the D2 cesium transition.  However, note that unless a single
cavity mode can be isolated, the heralded photon will contain more than one
cavity mode.  It is thus necessary to determine a minimum required cavity mode spectral
separation so that a single cavity mode may be isolated with existing spectral filters.
Using a fiber of length $L=5$cm leads [see Eq.~(\ref{sep})] to a spectral mode separation
of $\Delta \omega =13.5$GHz, which is many times the mode width $\delta \omega$ and could
therefore be isolated from the other modes.

The next step in our source design is to determine the required coefficient
of finesse, so as to ensure that the signal photon has a bandwidth which
matches that of the atomic transition.   Setting the signal-photon bandwidth to $2 \pi \times 5.22$MHz
in angular frequency, Eq.~(\ref{width}) determines the required coefficient
of finesse: $\mathscr{F}=6.2\times 10^4$ (or a corresponding reflectivity $r_2=0.992$).

\begin{figure}[ht]
\centering\includegraphics[width=0.4\textwidth]{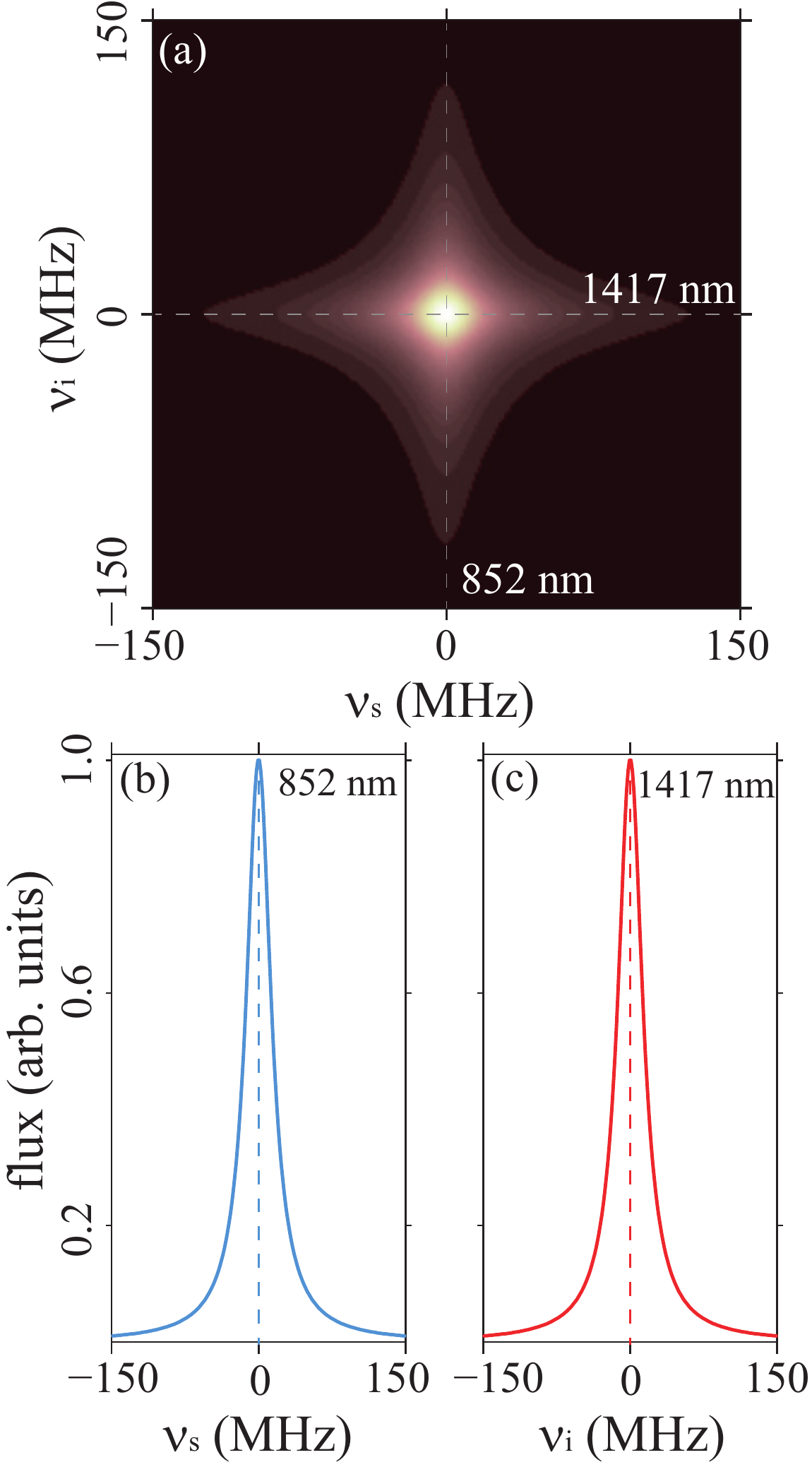}\caption{(a) Here we show
the joint spectral intensity, plotted as a function of signal and idler frequency detunings, obtained for a specific source design (see text), for which the signal photon has been matched in frequency and bandwidth to the D2 line transition of cesium.   (b) Marginal joint spectral distribution for the signal mode obtained by integrating the joint spectral intensity over the idler frequency.  (c) Marginal joint spectral distribution for the idler mode obtained by integrating the joint spectral intensity over the signal frequency.\label{Fig:diseSpec}}
\end{figure}

In Fig.~\ref{Fig:diseSpec}(a) we show a plot of the resulting joint spectral
intensity for our cavity source with the signal mode matched to the D2 line of cesium, where we have assumed a pump bandwidth of $\sigma=0.1$THz.  In panels (b) and (c) we show the corresponding single-photon spectra for the signal and idler modes.  Note
that in the next section we will describe a criterion for selecting the pump bandwidth.

\section{Photon-pair rate of emission}\label{SEC:flux}

In this section we focus on the emitted SFWM photon-pair flux for
a $\chi^{(3)}$ cavity. We define the
flux $N$ as the number of photons, say in the signal mode, emitted
from the cavity per second.  In the case of a pulsed pump, this corresponds
to the number of pairs emitted per pump pulse, multiplied by the pump
repetition rate $R$.  Thus, for a Csi cavity we obtain the following
expression

\begin{equation}
\label{flux1} N=R \sum_k
\langle\Psi_2^{si}|\hat{a}^{\dag}(k)\hat{a}(k)|\Psi_2^{si}\rangle.
\end{equation}

By substituting Eqs.~(\ref{zeta}) and (\ref{state3}) into Eq.~(\ref{flux1}),
and following a similar treatment as in Ref.~\cite{garay10}, which includes
writing sums over modes as integrals in the limit $\delta k \rightarrow 0$,
it can be shown that the emitted flux from a Csi cavity can be expressed as

\begin{eqnarray}
\label{flux2}
N=&&\frac{2^5c^2n^2(\omega_{o})
L^2\gamma^2p^2}{\pi^3\omega^2_{o}\sigma^2R}\nonumber\\&\times&\!\!\int\!\!\!
d\omega_s\!\!\int\! \!\!d\omega_i
\frac{\omega_s\omega_ik'(\omega_s)k'(\omega_i)}{n^2(\omega_s)n^2(\omega_i)}\mathscr{A}_{s}(\omega_{s})\mathscr{A}_{i}(\omega_{i})
\nonumber\\&\times&|f(\omega_s,\omega_i)|^2,
\end{eqnarray}

\noindent where $p$ is the average pump power,
$k'(\omega)=dk(\omega)/d\omega$, $\mathscr{A}_{\mu}(\omega_{\mu})$
is given by Eq.~(\ref{airy}), and
$f(\omega_s,\omega_i)=(\pi/2)^{1/2}\sigma F(\omega_s,\omega_i)$.
Note that this expression for the emitted flux
is similar to that obtained for SFWM in the absence of a cavity (see
Ref.~\cite{garay10}), except for the modified photon pair spectral
distribution due to the cavity mode structure.

For the monochromatic pump regime, the SFWM
emitted flux is obtained by taking the limit $\sigma\rightarrow 0$ of Eq.~(\ref{flux2}), from which we obtain

\begin{eqnarray}
\label{flux3}
N_{cw}=&&\frac{2^5c^2n^2(\omega_{p})L^2\gamma^2p^2}{\pi\omega^2_{p}}\nonumber\\&\times&\int
d\omega
\frac{\omega(2\omega_p-\omega)k'(\omega)k'(2\omega_p-\omega)}{n^2(\omega)n^2(2\omega_p-\omega)}\nonumber\\&\times&\mathscr{A}_{s}(\omega)\mathscr{A}_{i}(2\omega_p-\omega)\mbox{sinc}^2[L\Delta
k_{cw}/2],
\end{eqnarray}

\noindent where $\omega_p$ is the frequency of the
monochromatic pump field, and the phase mismatch $\Delta k_{cw}$ is
given by [see Ref.~\cite{garay10}]

\begin{eqnarray}
\Delta
k_{cw}=2k(\omega_p)-k(\omega)-k(2\omega_p-\omega)-2\gamma p.
\end{eqnarray}

\begin{figure}[ht]
\centering\includegraphics[width=0.4\textwidth]{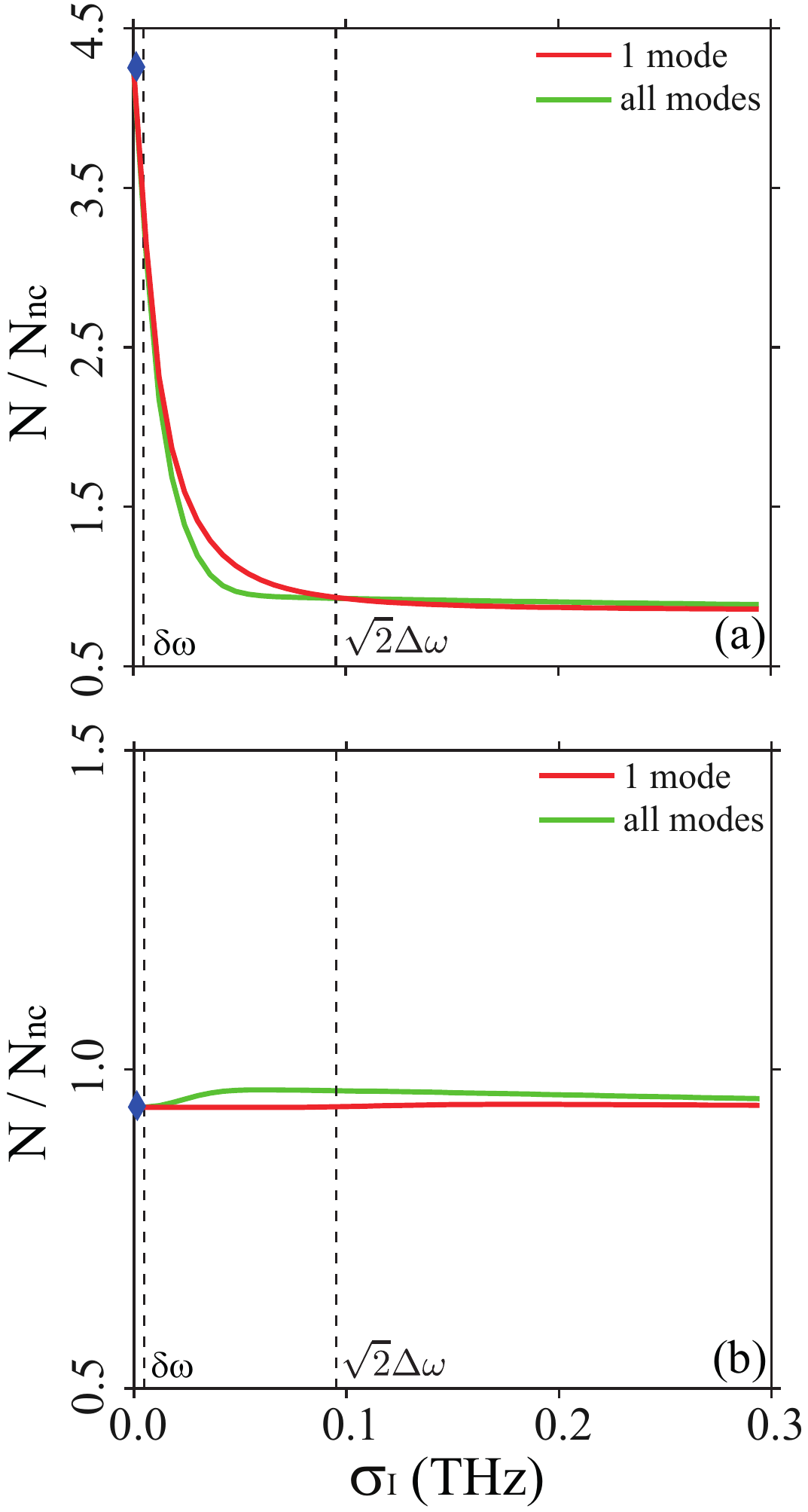}\caption{(a) Here we plot $N/N_{nc}$ versus the pump bandwidth $\sigma_I$, where $N$ is the SFWM photon-pair flux from a Csi cavity source and $N_{nc}$ is the flux of an equivalent source without a cavity.  Panel (b) is similar to (a), except that we have suppressed the idler-mode resonance (i.e. we have assumed a Cs cavity).
\label{Fig:Varsigma}}
\end{figure}

We are interested in a comparison of the SFWM flux obtained from a cavity source $N$, based on numerical integration of Eq.~(\ref{flux2}), to that obtained from an equivalent source
without a cavity, $N_{nc}$, also obtained from Eq.~(\ref{flux2}) with $\mathscr{F}_s=\mathscr{F}_i\rightarrow 0$.  Note that our numerical results below use the full two-photon state without any approximations.  For a numerical comparison, to be presented below, we assume a very similar source configuration as in our specific design presented in the previous section.  The main difference is that while in the specific design we assumed $r_2=0.992$ (or $\mathscr{F}=6.2\times 10^4)$, here we have assumed
$r_2=0.8$ (or $\mathscr{F}=80$).  While this lower coefficient of finesse facilitates our numerical calculations,
qualitatively the behavior is identical.

In Fig.~\ref{Fig:Varsigma}(a) we show plots of the quantity $N/N_{nc}$ as a function of the pump bandwidth $\sigma_I$, where $\sigma_I=\sqrt{2\ln(2)}\sigma$ is the intensity FWHM bandwidth of the pump, and while maintaining the average pump power constant. We have shown two curves, one corresponding to spectral filters of widths $5 \Delta \omega_s$ and
$5 \Delta \omega_i$ present on the paths of the signal and idler photons, so that a total of $25$ cavity modes are present in the two photon state (shown in green), and the other
corresponding to spectral filters $\Delta \omega_s$ and
$\Delta \omega_i$ so that a single cavity mode is present (shown in red).  It may be appreciated that there is a distinct behavior for each of the three zones: i) $\sigma_I> \sqrt{2} \Delta \omega$,  ii)  $\delta \omega<\sigma_I<\sqrt{2} \Delta \omega$, and iii) $\sigma_I<\delta \omega$.
In the first of these zones, $N/N_{nc}$ is independent of $\sigma_I$ with a value close to unity.   In other words, the flux produced by the cavity is essentially the same as that produced by an equivalent source without a cavity.   In the second zone, however, $N/N_{nc}$ rises as $\sigma$ is reduced until it reaches a maximum value within the third region.   The blue diamond shown in Fig.~\ref{Fig:Varsigma} indicates the flux calculated in the monochromatic pump limit, through numerical integration of Eq.~(\ref{flux3}).   This behavior implies that for $\sigma_I < \sqrt{2}\Delta \omega'$ there is an enhancement in the flux obtained
from the cavity, where the optimum enhancement occurs for small pump bandwidths values, within the region $\sigma_I < \delta \omega$.  Note that while the two curves drawn are qualitatively similar to each other, so that the flux enhancement effect occurs in both of these cases, because an atom-photon interface relies on single cavity mode, we are more interested in the flux enhancement obtained for a single mode.

Fig.~\ref{Fig:Varsigma}(b) is similar to Fig.~\ref{Fig:Varsigma}(a), computed for the Cs cavity instead of Csi cavity.  It is interesting to note that in this case $N/N_{nc}$ is close to unity throughout the pump bandwidth range considered.  In other words, there is no flux enhancement
observed for a SFWM cavity source which is resonant for only one of the two SFWM modes.  Thus, in order to observe a flux enhancement, the cavity needs to be resonant for both SFWM photons.

\subsection{Flux expressions obtained from a simple geometrical model}

The expression which we have derived for the SFWM flux, see Eq.~(\ref{flux2}), is in terms
of a two-dimensional integral.  While flux values may be obtained through numerical
integration of Eq.~(\ref{flux2}), as was done for Fig.~\ref{Fig:Varsigma}, we may gain additional physical insight from
a crude approximation of the flux
as $N=A H$ where $A$ is the area in $\{\omega_s,\omega_i\}$ space in which the joint spectrum
has an appreciable magnitude, and $H$ is the maximum value of the joint intensity, within this area.
For our present analysis we are interested in the SFWM flux obtained in a $\chi^{(3)}$ cavity, $N$, normalized
by the corresponding flux in an equivalent source without a cavity, likewise expressed as $N_{nc}=A_{nc} H_{nc}$.    The quantity $\xi$ may then
be approximated as $\xi \approx a d$, where $a\equiv H / H_{nc}$ and $d \equiv A / A_{nc}$.

Because the flux enhancement effect occurs for a single cavity mode in a manner qualitatively similar  as for multiple cavity modes (see Fig.~\ref{Fig:Varsigma}),
it is sufficient to analyze the former case in order to obtain an understanding
of the relevant physics.  Thus, in what follows we study the behavior of $\xi$ as a function
of the pump bandwidth, assuming that the
signal and idler photons are filtered using rectangular spectral filters of widths $\Delta \omega_s$ and
$\Delta \omega_i$, respectively, so that a single cavity mode is retained. Note that
while for non-degenerate SFWM, the intra-mode spacing $\Delta\omega$ and the mode width $\delta \omega$  differ between the two SFWM modes (for the specific design in the previous section this variation is near $5\%$), for the present analysis we will assume that $\Delta \omega_s=\Delta \omega_i$ and $\delta \omega_s=\delta \omega_i$.  We are interested in the three zones
i) $\sigma_I>\sqrt{2}\Delta \omega$,  ii)  $\delta \omega<\sigma_I<\sqrt{2}\Delta \omega$,
and iii) $\sigma_I<\delta \omega$ considered before.

Let us first analyze in this manner a Csi cavity.  In this case, it can be shown that parameter
$a$ is given by

\begin{equation}
a=\left( \frac{1+|r_2|}{1-|r_2|}\right)^2.
\end{equation}

\begin{figure}[ht]
\centering\includegraphics[width=0.4\textwidth]{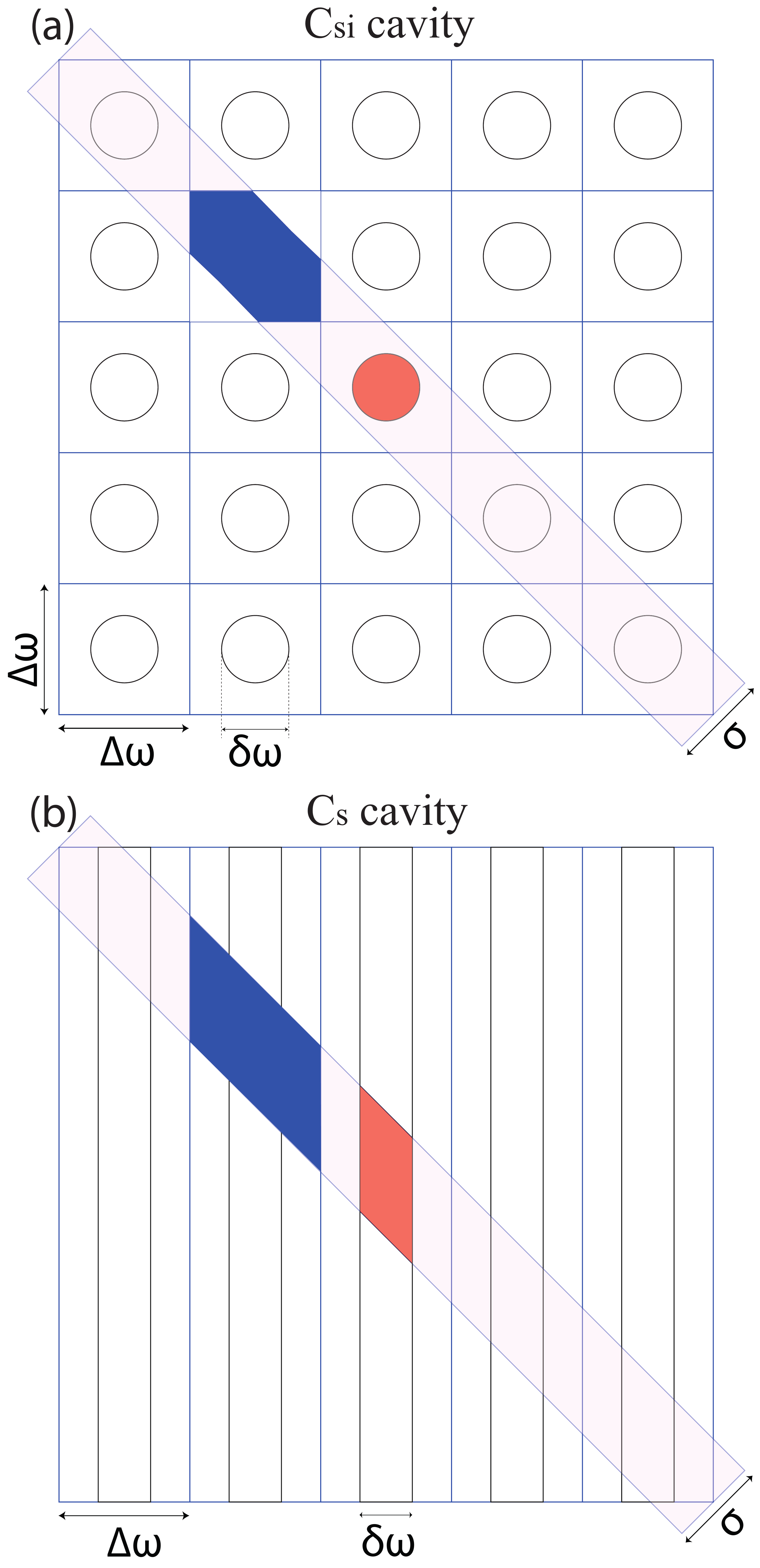}\caption{(a)  Here we show schematically the mode structure for
a Csi cavity in $\{\omega_s,\omega_i\}$ space.  Each circle, with diameter $\delta \omega$, represents a spectral cavity mode.  The square grid shown has a spacing given by the mode separation $\Delta \omega$, while the diagonal band, with width $\sigma_I$ indicates the pump spectral amplitude. (b) This panel shows a similar schematic, for a Cs cavity. The cavity modes are now elongated, with a width $\delta \omega$ for the signal mode.
\label{Fig:geometria}}
\end{figure}

In Fig.~\ref{Fig:geometria}(a) we show schematically the cavity mode structure. Each circle corresponds to a cavity mode of width $\delta \omega$, separated from each neighboring mode, along the signal and/or idler frequency axes, by $\Delta \omega$.   The spectral filters of width $\Delta \omega$ correspond to retaining only the flux produced within one of the squares in the grid shown.   The area shaded in red corresponds to the flux emitted with a cavity source, while the (square) area shaded in blue corresponds to the flux emitted with an equivalent source without a cavity.  

In region $1$, $A$ is the area of a disk of diameter $\delta \omega$, i.e. $A=\pi \delta \omega^2/4$, while $A_{nc}$ is the area of a square of dimension $\Delta \omega$, i.e. $ A_{nc}=\Delta \omega^2$, so that

\begin{equation}\label{ratio1}
\xi_1=a \frac{\pi \delta \omega^2}{4 \Delta \omega^2}=\frac{(1+|r_2|)^2}{4 \pi |r_2|},
\end{equation}

\noindent which is independent of $\sigma_I$.  In region $2$, $A=\pi \delta \omega^2/4$, while $A_{nc}=\sigma_I( \sqrt{2} \Delta \omega-\sigma_I/2)$, so that

\begin{equation}
\label{ratio2}
\xi_2=\frac{a\pi\delta\omega^2}{4\sigma_I(\sqrt{2}\Delta\omega-\sigma_I/2)}.
\end{equation}

In region $3$, we approximate $A$ as $A \approx \delta \omega \sigma_I$, an approximation which becomes better for smaller $\sigma_I$, while $A_{nc}=\sigma_I( \sqrt{2} \Delta \omega-\sigma_I/2)$, so that

\begin{equation}
\label{ratio3}\xi_3=\frac{a\delta\omega}{\sqrt{2}\Delta\omega-\sigma_I/2}.
\end{equation}

Note that the last two expressions for $\xi$ (Eqns.~\ref{ratio2} and \ref{ratio3}) depend on $\sigma_I$ and yield the flux enhancement described below. This leads to an expression for the flux enhancement, $E$, defined as the quotient  $\xi_1/\xi_3$, with $\sigma_I \rightarrow 0$.  We thus obtain

\begin{equation}
\label{incremento}E=\sqrt{2\mathscr{F}},
\end{equation}

\noindent making it clear that the flux enhancement increases with the coefficient of finesse.

Let us now turn our attention to a cavity source which
is resonant only
for one of the SFMW modes, i.e. a Cs or Ci cavity.  In this
case the parameter $a$ is given by

\begin{equation}
a=\frac{1+|r_2|}{1-|r_2|}.
\end{equation}

As can be seen in Fig.~\ref{Fig:geometria}(b), which is similar to Fig.~\ref{Fig:geometria}(a) drawn for the Cs cavity, in all three of the regions considered above, $A$ is the area of a parallelogram with height $\delta\omega$ and base $\sqrt{2}\sigma$, while $A_{nc}$ is also the area of a parallelogram with height $\Delta\omega$ and base $\sqrt{2}\sigma$. Thus, the ratio between the flux emitted from the cavity
and the flux emitted from SFWM without cavity is the same in the three regions and is
given by

\begin{equation}
\label{ratio4} \xi=\frac{a\delta\omega}{\Delta\omega}.
\end{equation}

This result indicates that in a Cs cavity there is no an enhancement
in the emitted flux due to the cavity, as was the case for a Csi cavity.
This behavior can be appreciated in Fig.~\ref{Fig:Varsigma}.

\subsection{Analysis of the absolute efficiency}

In our discussion of the SFWM flux,  we have so far focused on the flux attainable in a SFWM
cavity source, relative to an equivalent source without a cavity.  It is
likewise important to study the flux in absolute terms, which our analysis
in Sec.~\ref{SEC:flux} and in particular Eq.~(\ref{flux2}) permits.

While much of the behavior of the relative flux discussed so far is qualitatively identical to the behavior expected for a SPDC $\chi^{(2)}$ cavity source,
the absolute flux leads to some considerably different behaviors between the $\chi^{(2)}$ and $\chi^{(3)}$ cases.  This difference
is related to the fact that for SFWM, two pump photons are annihilated in each generation event, rather than just one for SPDC.  This implies that, within the phasematching bandwidth $\sigma_{PM}$ (i.e. for $\sigma_I \lesssim \sigma_{PM}$)  the SFWM flux scales linearly with the pump bandwidth $\sigma_I$, while the SPDC flux remains constant with respect to $\sigma_I$ for $\sigma_I \lesssim \sigma_{PM}$.  In physical terms, each pump frequency of one pump photon may participate in the SFWM process together with all pump frequencies of the other pump photon, limited by phasematching, so that increasing the pump bandwidth leads to a greater range of phasematched frequency combinations, and hence to a greater flux.   In contrast, in the case of SPDC, each pump frequency may be considered to act independently from other frequencies leading to a constant flux vs $\sigma$ dependence, for a constant average pump power.

\begin{figure}[ht]
\centering\includegraphics[width=0.4\textwidth]{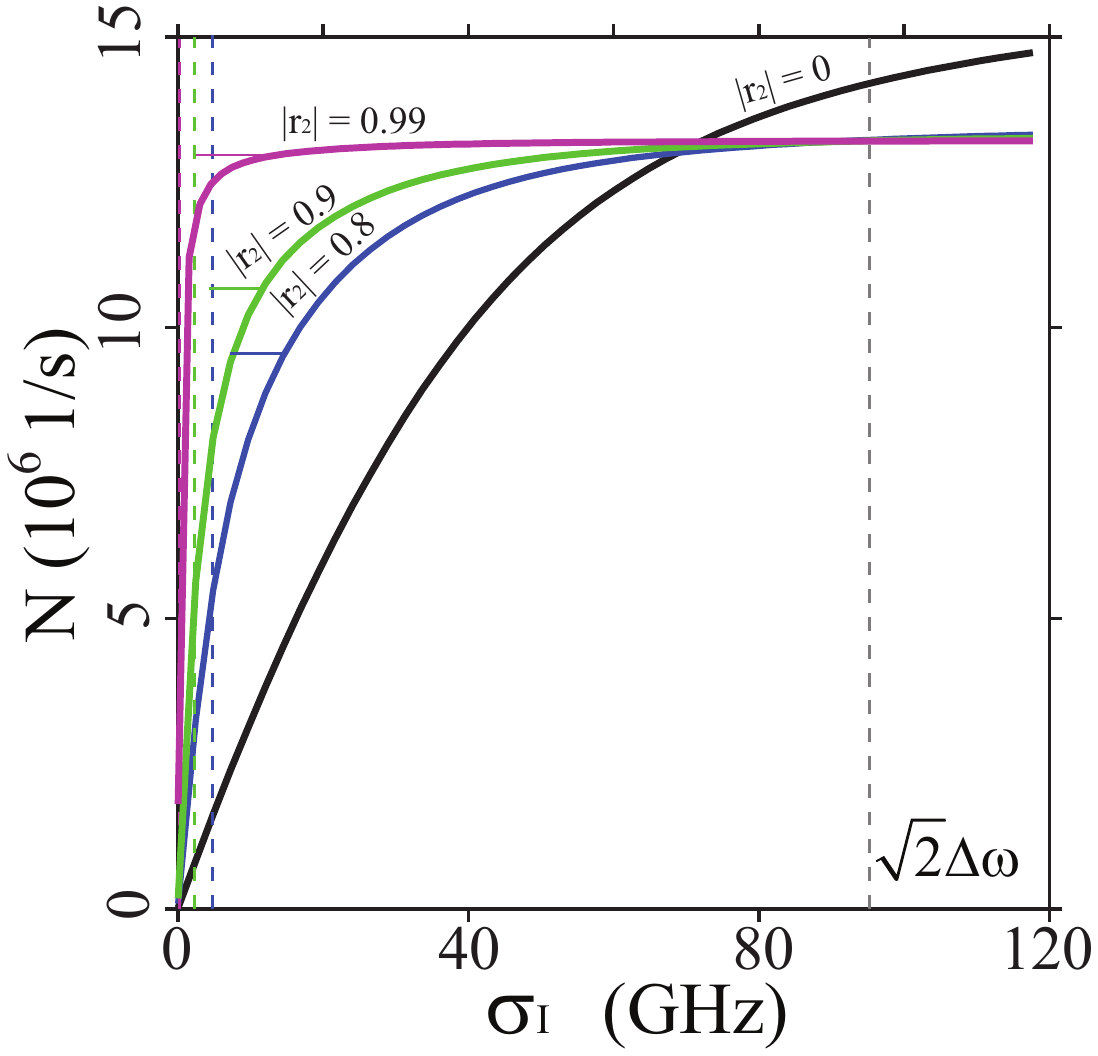}\caption{Here we show the absolute flux emitted by a Csi cavity  characterized by reflectivites $|r_2|=0.8,0.9,0.99$, with other source parameters selected as for Fig.~\ref{Fig:Varsigma}, as a function of the pump bandwidth $\sigma_I$. Also shown is the absolute flux as a function of $\sigma_I$ for an equivalent source without a cavity.  We have indicated with vertical dashed lines the $\delta \omega$ values obtained for each of these three cases, and we have also indicated the $\sqrt{2} \Delta \omega$ value.
\label{Fig:absflux}}
 \end{figure}

In Fig.~\ref{Fig:absflux}(a) we present the absolute SFWM flux, assuming a Csi cavity, plotted as a function of the pump bandwidth $\sigma_I$, for a number of different values of the reflectivity $|r_2|$, in particular for $|r_2|=0.8$, $|r_2|=0.9$ and $|r_2|=0.99$, with other source parameters selected to be identical to those assumed for Fig.~\ref{Fig:Varsigma}.   In this figure we have also indicated the cavity mode width  $\delta \omega$ value for each of these reflectivities and the $\sqrt{2} \Delta \omega$ value.  Note that the effect of the cavity is that the flux vs $\sigma_I$ dependence is linear for $\sigma_I \lesssim \delta \omega$, climbs more slowly for $ \sigma_I \gtrsim \sqrt{2} \delta \omega$, and thereafter reaches a plateau at a value of $\sigma_I$ which for large $\mathscr{F}$ is close to $\delta \omega$.    Thus, while the maximum cavity-induced flux enhancement occurs for $\sigma_I \rightarrow 0$, as is clear from Fig.~\ref{Fig:Varsigma}, the maximum absolute flux occurs within the above mentioned plateau.   For a large $\mathscr{F}$, the flux can be maximized if the pump bandwidth satisfies $\sigma_I \gtrsim \delta \omega$.   Note that in the case of Cs cavity, the absolute flux vs $\sigma_I$ curves (not shown) yield essentially the same curve as the case without cavity shown in Fig.~\ref{Fig:absflux}(a).

As a particular example, let us calculate the absolute flux for the specific source design of Section~\ref{Sec:SpDesign}.  This source involves degenerate pumps at $\lambda_p=1.064\mu$m, with SFWM wavelengths $\lambda_s=0.852\mu$m and $\lambda_i=1.417\mu$m.  We assume that the cavity source is based on a photonic crystal fiber of length $L=1$cm, with a core radius of $r=0.68\mu$m and air-filling fraction of $f=0.5$. We assume
a reflectivity of $|r_2|=0.998$, which corresponds to a coefficient of finesse of $\mathscr{F}=9.98\times10^5$, and a cavity mode width of $ \delta \omega=2\pi 5.22$ MHz, which defines resulting SFWM emission bandwidth and is matched to the D2 line transition of cesium. We assume an average pump power of $300$ mW,  with a  pump bandwidth $\sigma_I$ selected to have a value $\sigma_I=5 \delta \omega=0.164$GHz which fulfills the condition $\sigma_I \gtrsim \delta \omega$ of the previous paragraph. This value of $\sigma_I$ corresponds to a pulse duration of $16.9$ns, and we assume a repetition rate of $0.1$MHz. The resulting photon pair flux can then determined by numerical integration of Eq.~(\ref{flux2}): $1.03\times10^9$ photon pairs per second.

\section{Conclusions}

In this paper we have studied the generation of photon pairs through the process of spontaneous four wave mixing in $\chi^{(3)}$ cavities.  We have presented expressions for the two-photon state, showing that in the spectral domain the joint amplitude is given by the product of the corresponding joint amplitude in the absence of a cavity and a matrix of signal and idler modes defined by the cavity.  The width of each cavity mode $\delta \omega$ is proportional to $1/\sqrt{\mathscr{F}}$ where $\mathscr{F}$ is the coefficient of finesse.  Thus, the emission bandwidth can be effectively controlled by the quality of the cavity, which leads to the possibility of matching the frequency and bandwidth of one of the SFWM photons to a particular atomic transition, for the implementation of photon-atom interfaces.

We have also presented an analysis of the two-photon state in the temporal domain.   In particular, we have shown that the spectral cavity modes translate into a corresponding matrix of modes in the temporal domain.  The envelope which describes the amplitude of these modes has a non-factorable shape in the space formed by the times of emission.  This results in quantum entanglement in the temporal-mode degree of freedom with a dimensionality which scales with the effective number of cavity round-trips made by  the SFWM photons, as governed by the cavity finesse.  Also, we show that by approximating the cavity spectral modes through Gaussian functions it becomes possible to express the joint temporal intensity in closed analytic form.

We have presented expressions for the absolute source brightness, given in terms of a two-dimensional frequency integral.  We have evaluated this integral numerically to yield on the one hand a comparison of the flux obtained in the cavity to the flux obtained in an equivalent source without a cavity.  Through this analysis we have shown that the use of a pump bandwidth which is smaller than the inter-mode spacing results in a flux enhancement with respect to an equivalent source without a cavity.  We have presented a simple geometrical model which clarifies the physics behind the flux enhancement, and leads to an expression for the flux enhancement in terms of the coefficient of finesse.  Our analysis on the other hand shows that while the optimum source enhancement occurs for a small pump bandwidth $\sigma_I \rightarrow 0$, the absolute source brightness reaches its optimum value, in the case of a large coefficient of finesse, for $\sigma_I \gtrsim \delta \omega$, where $\delta \omega$ is the cavity mode width.  We hope that this work will be useful for the implementation of SFWM cavity sources, which could form a crucial component for atom-photon interfaces.

\section*{Acknowledgements}

This work was supported in part by CONACYT, Mexico,  by DGAPA, UNAM and by FONCICYT project 94142.


\end{document}